\def\fr#1#2{\hbox{${#1\over #2}$}}  

\def\ni{\noindent}       
\def\vs{\vskip.3cm}  
\def\Tr{\,{\rm Tr}\,}

\def\+{{(+)}}  \def\-{ {(-)} }   \def\0{ {(0)} }
\def\1{ {(1)} }  \def\2{ {(2)} }   
\def\pd{\partial}           

\def\m{\mu}              \def\n{\nu}              \def\k{\kappa}
\def\G{\Gamma}           \def\g{\gamma}           \def\d{\delta}
\def\S{\Sigma}           \def\s{\sigma}             
\def\a{\alpha}           \def\b{\beta}            \def\th{\theta}
\def\vphi{\varphi}       
\def\D{\Delta}                   
\def\con{\omega}          
\def\cs{\omega}
\def\ve{\varepsilon}           \def\O{\Omega}       
    \def\bO{\bar\Omega}
          \def\bv{{\bar v}}

\def\tf{\bar\varphi}

\def\d{\delta}
\documentstyle[11pt,eqs]{article}

\renewcommand{\theequation}{\thesection.\arabic{equation}}
\def\cleq{\setcounter{equation}{0}}
\def\nn{\nonumber} 
\def\be{\begin{equation}}             \def\ee{\end{equation}}
\def\ba{\begin{array}{rcl}}           \def\ea{\end{array}}
\def\beqa{\begin{eqnarray} }          \def\eeqa{\end{eqnarray} }
\def\beqalign{\begin{eqalign}}        \def\eeqalign{\end{eqalign}}
\def\leq#1{\label{eq:#1}}             \def\eq#1{(\ref{eq:#1})}
\def\bsubeq{\begin{subequations}}     \def\esubeq{\end{subequations}}
\def\bitem{\begin{itemize}}           \def\eitem{\end{itemize}}  
\begin{document}

\title{General Solution of the WZNW System   
          and 2D Induced Gravity in Curved Space-time\thanks{Work 
          supported in part by the Serbian 
          Science Foundation, Yugoslavia} }
\author{B. Sazdovi\'c\thanks{
       e-mail address: sazdovic@phy.bg.ac.yu}\\
       Institute of Physics, 11001 Belgrade, P.O.Box 57, Yugoslavia}       
\date{}
\maketitle
\begin{abstract} 
We find the general solution of the equations of motion for the 
WZNW system in curved space-time for arbitrary external gauge 
fields. Using the connection between the WZNW system for $SL(2,R)$ 
group and 2D induced gravity we obtain the general solution of 
the equations of motion for 2D induced gravity in curved 
space-time from that of the WZNW system. We independently presented 
the direct solution of 2D induced gravity equations of 
motion and obtain the same result.
\end{abstract}
\vs
\ni {\it PACS number(s)\/}: 04.60 Kz, 11.10 Kk, 04.20.Jb  \par
                                                        
\section{Introduction} 

It is well known that there exists a  connection between 2D induced gravity 
and $SL(2,R)$ WZNW theory which has been established in the light-cone 
gauge \cite{1,2,3} and in the conformal gauge \cite{4,5}. 
Recently the most complete connection  was found
where it has been shown that the 
$SL(2,R)$ WZNW system, defined as a difference of the two simple 
WZNW models, is gauge equivalent to the 2D induced gravity. The connection 
was established in a covariant way, using two independent approaches, 
lagrangian \cite{6} and hamiltonian \cite{7}. Both theories, 
WZNW system and 2D induced gravity, have full 
diffeomorphism invariance, which means that all previous results 
\cite{1,2,3,4,5} can be obtained  by partial gauge fixing of the action 
in refs. \cite{6,7}.

In ref.\cite{4} the reduction from WZNW theory to Liouville theory 
(2D induced gravity in the conformal gauge) was used to obtain the solution 
of Liouville equation of motion from the known solution of the WZNW 
equation of motion. As was pointed out by the authors, the advantage of 
this approach is that some configurations singular in Liouville 
variables, with regular energy-momentum densities, become regular in 
WZNW variables. We are going to generalize these results 
in the case of the curved space-time.  

In sec.2, for completeness of the paper, we introduce gauge invariant 
extension of the WZNW system, following ref. \cite{6}
and define the components of the currents and energy-momentum tensors.
Then, we solve the equations of motion for the WZNW system, in sec.3. Note, 
that this is a non trivial generalization of Witten's solution of the 
WZNW model, because we need the solution in the curved space-time for 
arbitrary external gauge fields. In sec.4 we again shortly repeat the result of 
ref.\cite{6}. We demonstrate how to make the transition from $SL(2,R)$ 
gauged WZNW system to 2D induced gravity, but this time with the help of 
the classical equations of motion, instead of functional integral method 
in ref.\cite{6}. Using the solution of the WZNW system, we find the 
general solution of equations of motion for 2D induced gravity in sec.5. 
In sec. 6 we present the direct solution of the equations of 
motion for 2D induced gravity.
The main results of the paper are discussed and summarized in sec. 7.
   
\section{WZNW system} 
\cleq

In this section we shall shortly repeat the consistent gauge invariant 
extension of the WZNW model, introduced in ref. \cite{6}.

The WZNW model is described by the action
\be
S(g)=S_0(v) + n\G(v) 
      =\fr{1}{2}\k\int_\S ({}^*v,v) + \fr{1}{3}\k\int_M (v,v^2)\, .
      \qquad               \leq{2.1}
\ee

The first term is an action of the non-linear $\s$--model 
for a group--valued field $g$: $\S \to G$ where $\S$ is a 
two--dimensional space time and $G$ is a 
semi-simple Lie group. The expression $(X,Y)=\fr{1}{2}\Tr(XY)$ is the
Cartan--Killing bilinear form, $v=g^{-1}dg\,$ is the Maurer--Cartan 
(Lie algebra valued) 1--form and ${}^*v$ is the dual of $v$. 
The second term is the topological Wess--Zumino term, defined on a 
three--manifold $M$ whose boundary is $\S=\pd M$. The constant $\k$ is 
defined as $\k =n \k_0$, where $n$ is an integer and   $\k_0$ is 
determined from the condition that the Wess-Zumino term is 
well defined modulo  $2\pi$. 

The action $S(g)$, is invariant under the {\it global\/} 
transformations    
\be
g\to g'=\O g\bO^{-1}\,                                 \leq{2.2}
\ee
where $\O$ and $\bO$ are  elements  of the group $G$.

To obtain the action invariant under the {\it local\/}
transformations we introduce two gauge fields $A$ and $B$ (Lie algebra 
valued 1--forms) and the {\it covariant derivative\/}:    
\be
Dg\equiv dg+Ag-gB \, ,\qquad (Dg)'=\O(Dg)\bO^{-1} \, .      \leq{2.3}
\ee
The covariant derivative $Dg$ transforms homogeneously under local
transformations, $(Dg)' $ $= \O (Dg) \bO^{-1}$, 
provided the gauge fields transform according to  
\be
A'=\O (A + d)\O^{-1} \, ,\qquad B'=\bO (B + d)\bO^{-1} \, . \leq{2.4}
\ee

Replacing 1--form $v=g^{-1}dg$ with the corresponding covariant 1--form  
\be
V\equiv g^{-1}Dg=v +g^{-1}Ag -B\, ,\qquad V'=\bO V\bO^{-1}\, ,\leq{2.5}
\ee
we can define the  gauge invariant extension of the WZNW action \eq{2.1}  
\be
S(g,A,B)=S_0(V)+n\G(V)
    =\fr{1}{2}\k\int_\S ({}^*V,V)+\fr{1}{3}\k\int_M (V,V^2)\, . \leq{2.6}
\ee

Although the action \eq{2.6} is gauge invariant, it is not defined as 
a field theory on a two-dimensional manifold $\S$ and it is not 
acceptable solution. It is useful to rewrite  the second term  in 
\eq{2.6} as   
\beqa
&&n\G(V)=n\G(v)+\G_1+\G_2+\G_3 \, ,\nn \\
&&\G_1=\k \int_\S \fr{1}{2}\Tr\bigl[
       -\bv A + vB +g^{-1}AgB \bigr] \, , \nn \\
&&\G_2=\k \int_M\fr{1}{2}\bigl[\cs_3(B,F_B)-\cs_3(A,F_A)\bigr]\, ,\nn \\
&&\G_3=\k \int_M \fr{1}{2}\Tr\bigl[
      F_{A}(Dg)g^{-1} +F_{B}g^{-1}(Dg) \bigr] \, ,        \label{2.7}
\eeqa
where $\cs_3(A,F_A)=\Tr(AF_{A}-\fr{1}{3}A^3)$ is the Chern--Simons
three--form. 
Our goal is to remove $\G_2$ and $\G_3$ from the action, because these forms 
are not exact and cannot be expressed as a integral over $\S$. We can simply 
omit $\G_3$ as a gauge invariant term, because the remainder is also 
gauge invariant. Reference  \cite{6} offers a new possibility for the solution 
of the $\G_2$ problem. Using the fact that $\G_2$  does not depend on $g$, the 
gauged WZNW system has been introduced as a difference of a two gauged 
WZNW models, so that the contribution of $\G_2$ terms canceled. In such a 
way the gauged WZNW system is just a difference of the regular parts of 
the action \eq{2.6} for two independent fields $g_1$ and $g_2$  
\bsubeq \label{2.8} 
\be
S(1,2)=S^r(g_1,A,B)-S^r(g_2,A,B) \, ,                    \label{2.8-a}
\ee
where the regular part is 
\be
S^r(g,A,B) =S(g) +\k \int_\S \fr{1}{2}\Tr\bigl[ 
   -{}^*\bv(A-{}^*A)+v(B+{}^*B) -\fr{1}{2}({}^*B +B)g^{-1}(A-{}^*A)g \bigr]\,
                                                        \label{2.8-b}
\ee
\esubeq 
with $\bv=gdg^{-1}$. 

It is useful to write the expressions for the metric tensor 
\be 
g_{\mu\nu} =e^{2F} {\hat g}_{\mu\nu}=
\fr{1}{2}e^{2F}\pmatrix{ -2h^-h^+    &  h^-+h^+ \cr
           h^-+h^+    &  -2      \cr }\, ,                \leq{2.9} 
\ee
and for the vectors in light-cone basis 
\be
U_{\pm}=e^{-F} {\hat U}_\pm =e_{\pm}{}^\mu U_\mu=
{\sqrt{2} e^{-F} \over h^- -h^+} (U_0+h^{\mp}U_1)\,  ,     \leq{2.10}
\ee
in terms of the light-cone variables $(h^+,h^-, F)$  \cite{6,7}.  
Then, the regular part of the action can be written as  
\be
S^r(g,A,B) =S(v) +\k \int_\S d^2\xi\sqrt{-g}\Tr\bigl[ 
   -\bv_-A_+ -v_+B_- -B_-g^{-1}A_+g \bigr]\,  ,           \leq{2.11}
\ee
where it is easy to check its invariance under the conformal rescaling 
$g_{\m\n}^\ve = e^{2 \ve} g_{\m\n}$, which implies $U_{\pm}^\ve =e^{-\ve}U_{\pm}$ 
for any vector $U_{\pm}$. Consequently, the field $F$ 
is absent and we will use ${\hat g}_{\m\n}$ eq. \eq{2.9} as 
a metric tensor, and ${\hat U}_{\pm}$ eq. \eq{2.10} as a vectors, 
but for simplicity from now we will omit the hat. 

Note that the action (\ref{2.8})  (or \eq{2.11}) depend only on the 
field combinations $A-{}^*A=2A_+ d\xi^+$ and $B+{}^*B=2B_-d\xi^-$, where 
$d\xi^{\pm}=e^{\pm}{}_\mu d\xi^\mu$.

Because the components $A_-$ and $B_+$ are effectively absent 
we can write covariant derivatives in the forms 
\be
D_+g=\pd_+g+A_+g \qquad D_-=\pd_-g-gB_-  \,  .        \leq{2.12}             
\ee

Therefore, the independent fields in the theory are:
two matrix-valued fields $g_1$ and $g_2$, components of gauge fields 
$A_+$ and $B_-$ and two components of the metric tensor $h^+$ and $h^-$ .

For both regular parts of the action we define the components of the currents 
\beqa
&&J_{i-}=-{\d S^r (g_i) \over 2\k \sqrt{-g} \d A_+}=-(D_-g_i)g_i^{-1}\, , \nn\\  
&&J_{i+}=-{\d S^r(g_i) \over 2\k \sqrt{-g} \d B_-}=g_i^{-1}D_+g_i \, , \qquad 
 (i=1,2)                                          \label{2.13}                 
\eeqa
and the components of the energy-momentum tensors 
\be
T_{i\pm}=\pm 4 {\d S^r(g_i) \over \k \d h^{\pm}}\,  . 
  \qquad  (i=1,2)                                   \leq{2.14} 
\ee

Later, we will need the connection 
\be
T_{i+}=Tr(J_{i+}^2-A_+^2), \qquad  T_{i-}=Tr(J_{i-}^2-B_-^2) \, ,  \leq{2.15}           
\ee
which can be verified using the expression ${\d \over \d h^{\pm}}\int 
Tr({}^*AB)=\pm \int d^2\xi Tr(A_{\pm}B_{\pm})$.

\section{General solution of the equations of motion for WZNW system} 
\cleq

Variation of the action (\ref{2.8}) with respect to the independent 
variables $g_i^{-1}\d g_i$ \, $(i=1,2)$, $\d A_+$, 
$\d B_-$ and $\d h^{\pm}$  give us the following equations 
of motion respectively
\bsubeq \label{3.1}
\be
F_i\equiv d W_i+W_i^2=0 \, ; \qquad W_i \equiv J_{i+} d\xi^+ +B_-d\xi^- \, 
\quad (i=1,2) ,                             \label{3.1-a}
\ee
\be
J_{1\mp}=J_{2\mp} \, ,                     \label{3.1-b}
\ee
\be
T_{1\pm}=T_{2\pm}  \, .                    \label{3.1-c}
\ee
\esubeq
In derivation of the equations (\ref{3.1-a}) we used the fact that the 
self-dual part $B+{}^*B$ as well as the anti-self-dual part $J-{}^*J$ are 
nilpotent.

The equations (\ref{3.1-b}) show that the currents of the sectors 1 and 2 are equal 
and we can simply omit the index $i$, preserving the same expressions 
$J_{\pm}$, for both currents.
As a consequence of this and the relations \eq{2.15}  
the equations (\ref{3.1-c}) are automatically fulfilled. 

For the same reason the equations (\ref{3.1-a}) are 
equivalent and we should solve only one equation $F\equiv dW+W^2=0$, with  
$W \equiv  J_+d\xi^++B_-d\xi^-$. Its solution is just {\it pure gauge} 
because the field strength is equal to zero. Since under gauge transformations 
the $W$ field transforms as $W'=\bar \O (W+d){\bar \O}^{-1}$, 
we have  
\be
W=\bar \O d {\bar \O}^{-1}\,  ,                     \leq{3.2}
\ee
or in component notation $J_+={\bar \O} \pd_+ {\bar \O}^{-1}$ and 
$B_-=\bar \O \pd_-  {\bar \O}^{-1}$. After some rearrangement we can write 
these equations in the form $D_+(g \bar \O)=0$ and $D_-({\bar \O}^{-1})=0$.    

Let us introduce $G_{\mp}$, as a group-valued functions annihilated by the 
covariant derivatives $D_{\pm}$ 
\be
D_{\pm} G_{\mp} =0 \,  .                             \leq{3.3}
\ee
Then we can put $g \bar \O =G_-$,  ${\bar \O}^{-1} =G_+$ , and after 
eliminating the parameter $\bar \O$ we obtain 
\be
g=G_- G_+ \,  .                                      \leq{3.4}
\ee
Consequently, the equation \eq{3.4} with constraints \eq{3.3} is the 
solution of the equation of motion (\ref{3.1-a}).

Let us record, that on this solution we have 
\be
J_+=G_+^{-1} \pd_+ G_+  \, , \qquad    J_-=G_- \pd_- G_-^{-1} \,  ,  \leq{3.5}
\ee and from constraints \eq{3.3} we can find 
\be
A_+= G_- \pd_+ G_-^{-1} \, , \qquad  B_-=G_+^{-1} \pd_- G_+ \, . \leq{3.6}
\ee

It remains to solve the equation (\ref{3.1-b}). With the help of \eq{3.5} 
we have 
\be
G_{2+} =e_+ G_{1+} \,  ,  \qquad   G_{2-}=G_{1-} e_-        \leq{3.7}
\ee
where $e_{\pm}  \in G$, 
and $\pd_{\pm} e_{\pm}=0$. Applying covariant derivatives $D_-$ and $D_+$ on the  
eqs. \eq{3.7} respectively and using \eq{3.3} we get $\pd_{\mp} e_{\pm} =0$ 
and consequently $e_{\pm} =$ const. 

Therefore, our solution of the equations of motion (\ref{3.1}) is 
\be
g_1 = G_- G_+  \,  , \qquad   g_2 = G_- e G_+  \,   , 
\qquad (D_{\pm} G_{\mp} =0)                                     \leq{3.8}                                     
\ee
where we introduce $G_{\pm}$ instead of $G_{1\pm}$, $G_\pm \equiv G_{1\pm}$ 
and a new constant matrix $e=e_- e_+ \in G$ .

Let us conclude that for arbitrary given functions $G_{\pm}$ we can find 
the matrix valued fields $g_1$ 
 and $g_2$ from \eq{3.8} as well as the gauge fields $A_+$ and $B_-$ from 
\eq{3.6}. On the other hand, if we have arbitrary external fields $A_+$ and 
$B_-$, solving constraints \eq{3.3} we can find $G_\mp$ and then  
$g_1$ and $g_2$ with the help of \eq{3.8}. In both cases the metric tensor 
$g_{\mu \nu}$ is arbitrary.   

The solution for one sector can be expressed in a different way. If we 
represent the components of the vector fields $A_+$ and $B_-$ , with 
corresponding matrices $\a$ and $\b$ belonging to the group $G$, as 
\be
A_+= \a \pd_+ \a^{-1} \, , \qquad   B_-= \b^{-1} \pd_- \b \, ,   \leq{3.9}
\ee
we can write the conditions \eq{3.3} as $\pd_+ (\a^{-1} G_- )=0$  and 
$\pd_- (G_+ \b^{-1} )=0$ . In that case the solution of the equation 
(\ref{3.1-a}), instead of \eq{3.3} and \eq{3.4} can be written in the form 
\be
g=\a g_- g_+ \b \,   .  \qquad     (\pd_{\pm} g_{\mp} =0)     \leq{3.10}
\ee

It is instructive to consider some particular cases. In the absence of 
external fields $A_+ =0= B_-$, we have
\be
g=g_- g_+ \,   .   \qquad     (\pd_{\pm} g_{\mp} =0)     \leq{3.11}
\ee

For flat space-time, $h^{\pm}=\mp 1$ we have 
$\pd_{\pm} = {1 \over \sqrt{2} }(\pd_0 \pm \pd_1)$ and we can 
solve the constraints $\pd_{\pm} g_{\mp}=0$, finding  
$g_{\pm} =g_{\pm} (\xi^0 \pm \xi^1)$ . Therefore, in   
flat space-time and in the absence of external fields we obtain the 
well known Witten solution \cite{8} 
\be
g=g_- (\xi^0 - \xi^1 ) g_+ (\xi^0 + \xi^1 )    \leq{3.12}
\ee
of the equation of motion for one WZNW model.

Note that because $W_+=J_+$ and $W_-=B_-$  from \eq{3.5} and \eq{3.6} we have 
$W=G_+^{-1} d G_+$ and the equation (\ref{3.1-a}) is just a Maurer-Cartan 
equation. Similarly, if we introduce ${\bar W}_+=A_+$ and ${\bar W}_-=J_-$ 
we have ${\bar W} =G_- d G_-^{-1}$  and $\bar F =d \bar W + {\bar W}^2 =0$ . 
This is also the equation of motion for the variation of the action (\ref{2.8})   
but this time with respect to $g \d g^{-1}$ .

\section{Transition from WZNW system to 2D induced gravity}  
\cleq

It has been shown in ref. \cite{6}, that if we take a group $H=H_+ \times H_-$ 
($H_+$ and $H_-$ are subgroups of $SL(2,R)$ defined by the generators 
($t_+ , t_0$ ) and ($t_0 , t_-$) respectively), instead of working with the 
arbitrary gauge group $G \times G$, and  integrating out some 
variables we obtain the induced gravity action. In this section we shall 
demonstrate the same result using classical equations of motion. 

First we take $G=SL(2,R)$ and use the Gauss decomposition in the  
neighborhood of the identity 
\be
g = e^{x t_{(+)}} e^{\vphi t_\0} e^{y t_{\-}} 
  = e^{-\vphi/2}  \pmatrix{ e^{\vphi}+xy   &  x  \cr
                                y        &  1  \cr }\, .    \leq{4.1}
\ee

The generators of the group $SL(2,R)$ are 
$t_{(\pm)}={\fr 1 2}(\s_1\pm i\s_2)$ and  $t_\0 ={\fr 1 2}\s_3$,  
the nonzero elements of the Cartan metric are $\g_{+-}=\g_{-+}=2$ and  
$\g_{00}=1$, and the nontrivial structure constant is $f_{+-0}=2$. 

To select the group $H_+ \times H_-$ for the gauge group, it is equivalent 
to first take the group $SL(2,R) \times SL(2,R)$ and then restrict it as 
\be
A_+^\-=0 \,  , \qquad  B_-^{(+)} =0 \,   .                \leq{4.2}
\ee
In this case the independent fields are $g_1 \to q_1^\a =(x_1, \vphi_1, y_1)$,  
$g_2 \to q_2^\a =(x_2, \vphi_2, y_2)$, $A_+^{(+)} \, , A_+^\0\, , B_-^\0\, 
B_-^\-\, ,h^+\,$ and $h^-$. In terms of these variables, we can write the action in the 
form
\bsubeq \label{4.3} 
\be
S(1,2)=S^r (1) - S^r (2) \, ,  \label{4.3-a}
\ee
with the regular part  
\beqa
S^r(x,\vphi,y,A,B) = \k\int d^2\xi\sqrt{-g}\,&&\Bigl[ 
\pd_-\vphi\pd_+\vphi+2A^\0_{+}\pd_-\vphi -2B^\0_{-}\pd_+\vphi\nn\\
&&\hskip30pt +4D_+xD_-y\,e^{-\vphi} -2B^\0_{-}A^\0_{+} \Bigr] \, , \label{4.3-b}
\eeqa
\esubeq  
where the covariant derivatives on the group manifold are 
\be
D_+ x =\bigl[\pd_+ +A^\0_+\bigr]x +A^{(+)}_+ \, ,\qquad
D_- y =\bigl[\pd_- -B^\0_-\bigr]y -B^\-_-   \, ,   \leq{4.4}
\ee
and $\sqrt{-g} = \fr{1}{2} (h^- - h^+)$ in agreement with \eq{2.9}.

In component notation, instead of the eq. (\ref{3.1-a}) 
we have three equations of motion  
\bsubeq \label{4.5}
\be
(\nabla_+ - A_+^\0 ) J_-^\- =0 \,  ,    \label{4.5-a} 
\ee
\be
\Delta \vphi -2 \nabla_- A_+^\0 +2 \nabla_+ B_-^\0 +4 J_+^{(+)} 
J_-^\- e^{\vphi}=0 \,  ,        \label{4.5-b}  
\ee
\be
(\nabla_- + B_-^\0 ) J_+^{(+)} =0 \, ,          \label{4.5-c}
\ee
\esubeq
which can be also obtained varying the action (\ref{4.3}) with respect to fields 
$x, \vphi, y$ respectively.
Here 
\be
\nabla_{\pm} X_n = (\pd_{\pm} +n \con_{\pm}) X_n           \leq{4.6}
\ee 
are covariant derivatives on tensor $X_n$ (the number $n$ is sum of the 
indices, counting index $+$ with $1$ and index $-$ with $-1$), 
$\con_{\pm}$ are Riemannian connections and 
\be
\Delta = -2 \nabla_- \pd_+ =-2 \nabla_+ \pd_-             \leq{4.7}
\ee
is the Laplace operator. From equations (\ref{4.5-a}) and (\ref{4.5-c})
we obtain 
\be
A_+^\0 =-\con_+ + \pd_+ \ln \vert J_-^\- \vert \,  , \qquad 
B_-^\0 =-\con_- - \pd_- \ln \vert J_+^{(+)} \vert \,  .    \leq{4.8}
\ee
Substituting this in the (\ref{4.5-b})  we get 
\be
\Delta \tf +R+\m e^{\tf} =0 \,  ,          \leq{4.9}
\ee
where 
\be
R =2 \nabla_- \con_+ -2 \nabla_+ \con_-                  \leq{4.10}
\ee
is scalar curvature and we introduced a new variable $\tf$, defined by the equation   
\be
\m e^{\tf} = 4J_+^{(+)} J_-^\- e^{\vphi} \,  .               \leq{4.11}  
\ee
Note that the essential role of the constant $\m$ is the sign of 
$J_+^{(+)} J_-^{\-}$, while its amplitude can easily be set equal to one by 
rescaling the field $\tf$. We take $\m > 0$ because in this case the action 
is bounded from below. 

The equation of motion \eq{4.9} can be obtained varying the  regular part 
of the action 
\be
S^r( \tf) =\k \int d^2 \xi \sqrt{-g} 
[\fr{1}{2} \tf \D \tf + \tf R +\m e^{\tf} ]                \leq{4.12}
\ee   
with respect to $\tf$, and consequently the action 
\be
S=S^r( {\tf}_1)-S^r( {\tf}_2) =  \int d^2 \xi \sqrt{-\bar g} 
[\fr{1}{2} \phi {\bar \D} \phi + \fr{\a}{2} \phi {\bar R} + 
 M (e^{\fr{2 \phi}{\a}}-1) ] \, , \qquad (M=\k \m )          \leq{4.13}
\ee   
is equivalent to (\ref{4.3}) on the equations of motion,  
where we have introduced new variables 
\be 
\phi = \fr{\a}{2} ({\tf}_1 - {\tf}_2 ) \, , \qquad 
{\bar g}_{\m\n} = e^{{\tf}_2} g_{\m\n} \,  . 
\qquad   (\a=2 \sqrt{\k})                                 \leq{4.14}                                      
\ee
Under conformal rescaling introduced in sec.2., we have 
$g_{\m\n}^{\ve} =e^{2\ve} g_{\m\n}$ and with the help of \eq{4.11} we find 
$\tf^{\ve}=\tf -2 \ve$. Therefore we can interpret the new variables as the 
conformaly gauge invariant variables.

The action \eq{4.13} is well known as 2D induced gravity action. The 
same result, equivalence 
between WZNW system (\ref{2.8}) for the group $H=H_+ \times  H_-$ and induced gravity 
\eq{4.13} was  established in  ref.\cite{6} using path integral methods.

Note, that for transition from WZNW system to 2D induced gravity we essentially 
chose the gauge group $G=SL(2,R)$ and than put the constraints \eq{4.2}.

\section{General solution of the equations of motion for 2D induced gravity 
from that of WZNW system} 
\cleq

Using the solution of the equations of motion for WZNW system 
\eq{3.6} and  \eq{3.8} we can find the solution of the equations of 
motion for 2D induced gravity with the help of transition condition established 
in the previous section.

Because the functions $G_{\pm}$, as well as the field $g$, are elements 
of the $SL(2,R)$ group we will use for them a similar Gauss 
decompositions  as  \eq{4.1}
\be
G_{\pm} =  e^{-G_{\vphi \pm}/2} 
\pmatrix{ e^{G_{\vphi \pm}}+G_{x \pm} G_{y \pm} & G_{x \pm}  \cr
                              G_{y \pm}         & 1   \cr }\, . \leq{5.1}
\ee
For constant matrix $e \in SL(2,R)$ we simply put 
\be
e= \pmatrix {a   &  b   \cr 
             c   &  d   \cr }\, .  \qquad  (ad-bc=1)    \leq{5.2}
\ee

From eqs. \eq{3.6} and \eq{5.1} we find   
$A_+^\- =-e^{-G_{\vphi -}} \pd_+ G_{y -}$ and 
$B_-^{(+)}=e^{-G_{\vphi +}} \pd_- G_{x +}$, and consequently the 
constraints \eq{4.2} give us  
\be
\pd_+ G_{y -} =0 \,  , \qquad  \pd_- G_{x +} =0 \,  .        \leq{5.3}
\ee

Now, using decomposition \eq{5.1}, \eq{5.2} and simple matrix multiplication, 
we obtain from the solution of the equations of motion for WZNW system \eq{3.8} 
\be
e^{\vphi_1} = {e^{G_{\vphi +}+G_{\vphi -}} \over (1+ G_{x +}G_{y -})^2 } 
                                                          \leq{5.4}
\ee
and 
\be
e^{\vphi_2} = {e^{G_{\vphi +}+G_{\vphi -}} \over 
(d+ c G_{x +}+ b G_{y -}+a G_{x +}G_{y -})^2 } \,  .     \leq{5.5}
\ee

In fact we need the solution for fields $\tf_i$, which can be easily found 
with the help of the definitions \eq{4.11}  
\be
e^{\tf_1} = {4 \pd_+ X_+ \pd_- X_- \over \m (1- X_+ X_-)^2 } \,  , \qquad
e^{\tf_2} = {4 \pd_+ Y_+ \pd_- Y_- \over \m (1- Y_+ Y_-)^2 } \,   , \leq{5.6}
\ee
where we introduced new notation $X_+ \equiv G_{x +}$, $X_- \equiv 
-G_{y -}$, $Y_+ =a X_+ +b \,$ and $Y_- =a X_- -c \,$. In  derivation 
of \eq{5.6} we have used the expression for the currents 
$J_+^{(+)} =e^{-G_{\vphi +}} \pd_+ G_{x +}$ and 
$J_-^{(-)} =-e^{-G_{\vphi -}} \pd_- G_{y -}$ obtained from \eq{3.5} and 
\eq{5.1}.  

Finally, the solution for induced gravity fields $\phi$ and 
${\bar g}_{\m\n}$ introduced in \eq{4.14} is 
\beqa
&&\phi = \a \ln{1-Y_+ Y_- \over a(1-X_+ X_-)} \, ,   \nn\\ 
&&{\bar g}_{\m\n} ={\a^2 \over 2M}{\pd_+ Y_+ \pd_- Y_- \over (1- Y_+ Y_-)^2 } 
 \pmatrix {-2 h^- h^+    &  h^- + h^+   \cr 
             h^- + h^+   &  -2          \cr }\, ,   \label{5.7}
\eeqa
where, as before, $Y_+ =aX_+ +b ,  \quad Y_-=aX_- -c \,$ and $\pd_{\pm} X_{\mp}=0 $ 
as a consequence of \eq{5.3}. 

After some calculations we can check that the scalar curvature 
corresponding to the metric tensor ${\bar g}_{\m\n}$ in (\ref{5.7}) 
is constant ${\bar R} = -\m$. 

Therefore, in the solution of 2D induced gravity we have in fact a 
constant curvature metric tensor ${\bar g}_{\m\n}$, two arbitrary 
functions $X_+ (\eta_+)$ and $X_- (\eta_-)$, depending on the fields  
$\eta_{\pm}$ which are solutions of the conditions $\pd_{\pm} \eta_{\mp} =0$, 
and three arbitrary constants $a,b$ and $c$. 

The same result can be explained in a different way. We can say that we have 
two arbitrary fields $\eta_+$ and $\eta_- $, two arbitrary functions 
$X_{\pm}(\eta_{\pm})$ and three arbitrary constants $a,b$ and $c$. In this  
case 
\be
h^{\pm} =- {\pd_0 \eta_{\pm} \over \pd_1 \eta_{\pm} } \,  ,    \leq{5.8}
\ee
so that conditions $\pd_{\pm} \eta_{\mp}=0$ are satisfied.

\section{Direct solution of the equations of motion for 2D induced gravity} 
\cleq

It is instructive to solve the equations of motion for 2D induced gravity 
\eq{4.13} directly. The equations of motion for 
fields $\phi$ and the conformal factor of ${\bar g}_{\m\n}$, $\tf_2$ 
respectively are  
\bsubeq \label{6.1}
\be
{\bar \Delta}\phi +\fr{\a}{2} {\bar R} + {2M \over \a} 
e^{2\phi \over \a}=0 \,  ,                                 \label{6.1-a}
\ee
\be
\fr{\a}{2}{\bar \Delta}\phi +M (e^{2\phi \over \a}-1)=0 \,  .  \label{6.1-b}
\ee
\esubeq
These equations imply that ${\bar g}_{\m\n}$ has constant curvature 
${\bar R}=- \fr{4M}{\a^2}=-\m$, as we already obtained from solution 
(\ref{5.7}). Then, from the identity $\sqrt{-{\bar g}}{\bar R} = 
\sqrt{-g}(R+ \Delta \tf_2)$ we have 
\bsubeq \label{6.2}
\be
\Delta \tf_2 +R+ \m e^{\tf_2} =0 \,  ,                        \label{6.2-a}
\ee
and from (\ref{6.1-b}) we find 
\be
\Delta (\fr{2\phi}{\a} +\tf_2 ) +R+ \m e^{\fr{2\phi}{\a} +\tf_2} =0 \, . 
 \label{6.2-b}
\ee
\esubeq
This means that the fields $\fr{2\phi}{\a} +\tf_2$ and $\tf_2$ satisfy 
the same equations, which are in fact eqs. \eq{4.9}, because we know from 
\eq{4.14} that $\fr{2\phi}{\a} +\tf_2 = \tf_1$. 
We prefer to use $\tf_1 \, , \tf_2 \, , h^+$ and 
$h^-$ as an independent variables, instead of $\phi$ and ${\bar g}_{\m\n}$. 
 
We already know that the solutions of eqs. (\ref{6.2}) have the form as in   
\eq{5.6} but here we put some independent functions $X_{1 \pm}$ and $X_{2 \pm}$ 
instead of $X_{\pm}$ and $Y_{\pm}$
\be
e^{\tf_i} = {4 \pd_+ X_{i+} \pd_- X_{i-} \over \m (1- X_{i+} X_{i-})^2 } \, . 
\qquad  (i=1,2)                                       \leq{6.3}
\ee

The equations of motion for the fields $h^{\pm}$ are 
\be
\Theta_{1 \pm} = \Theta_{2 \pm} \,  ,                \leq{6.4}  
\ee
where 
\be
\Theta_{i \pm} = \mp {\d S^r (\vphi_i ,h^+,h^-) \over \k \d h^{\pm} } 
    \qquad (i=1,2)                                   \leq{6.5}
\ee
are the components of the energy-momentum tensors for 2D induced gravity. 
In this case we have 
\be
\Theta_{\pm} = -\fr{1}{2}(\pd_{\pm} \tf )^2 \mp  \nabla_- \pd_{\pm} \tf \pm  
\nabla_+ \pd_{\pm} \tf +\fr{\m}{2}e^{\tf} \, ,           \leq{6.6}
\ee
or with help of (\ref{6.2}) 
\be
\Theta_{\pm} =  \nabla_{\pm} \pd_{\pm} \tf -\fr{1}{2}(\pd_{\pm} \tf)^2 
-\fr{1}{2} R \, .                                         \leq{6.7} 
\ee
On the solution \eq{6.3} we can write 
\be
\Theta_{\pm} =  \{ X_{\pm} , \xi \} - \fr{1}{2} R          \leq{6.8}
\ee
where 
\be
\nabla_{\pm} \pd_{\pm} \tf -\fr{1}{2}(\pd_{\pm} \tf)^2  =
{ \nabla_{\pm}^3 X_{\pm}  \over \pd_{\pm} X_{\pm}} 
-\fr{3}{2} \Bigl( { \nabla_{\pm}^2 X_{\pm}  \over \pd_{\pm} X_{\pm}} {\Bigr)}^2  
\equiv \{X_{\pm} ,\xi \}                                  \leq{6.9}
\ee
are  Schwarzian  derivatives. Because only$X_{i \pm}$ variables depend on the 
sectors 1 and 2 the equation \eq{6.4} simply mean the equalities of the 
Schwarzian derivatives  
\be
\{X_{1 \pm} , \xi \} =\{X_{2 \pm} , \xi \} \,  .        \leq{6.10}
\ee
With help of the identity 
$\{X_{2 \pm} , \xi \} =
\{X_{2 \pm}, X_{1 \pm} \} (\pd_{\pm} X_{\pm})^2 + \{X_{1 \pm} ,\xi \}$
we obtain  
\be
\{X_{2 \pm}, X_{1 \pm} \} =0 \,  ,                    \leq{6.11}
\ee
which leads to the connection between $X_{1 \pm}$ and $X_{2 \pm}$  
\be
X_{2 \pm} = { a_{\pm} X_{1 \pm} + b_{\pm}  \over 
c_{\pm} X_{1 \pm} + d_{\pm} } \,  , \qquad              \leq{6.12}
\ee
where $a_{\pm}, b_{\pm}, c_{\pm}$ and $d_{\pm}$ are constants satisfying 
the condition $a_{\pm} d_{\pm} - b_{\pm} c_{\pm} =1$. 

Here we can see the advantage of the WZNW variables. The equation \eq{6.12} 
can be obtained from \eq{3.7} using matrix multiplication if we take 
\be
e_+= \pmatrix {a_+   &  b_+   \cr 
               c_+   &  d_+   \cr }\, , \qquad
e_-= \pmatrix {a_-   & - c_-   \cr 
               -b_-  &  d_-   \cr }\, ,                \leq{6.13}
\ee
and use decompositions \eq{5.1} and definitions of $X_{\pm}$ . 

Substituting \eq{6.12} into \eq{6.3} we obtain eqs. \eq{5.6} where 
$X_{\pm}\equiv X_{1\pm}$, $Y_+=aX_+ +b$, $Y_-=aX_- -c$ and 
$a ,b$ and $c$ can be obtained in terms of     
$a_{\pm}, b_{\pm}, c_{\pm}$ and $d_{\pm}$ with the help of \eq{5.2} and 
 the relation $e=e_- e_+$. The final solution (\ref{5.7}) is the same as in sec. 5.

\section{Discussion and conclusions} 
\cleq

In this paper we solved the equations of motion for the WZNW system and 
2D induced gravity in curved space-time.

First we solved the equations of motion for the WZNW system in arbitrary 
external fields and for curved backgrounds. We found that this solution depends 
on the two arbitrary matrix functions $G_{\pm}$ and one constant matrix 
$e$, all belonging to the group $G$. The metric tensor is arbitrary. 

We presented two independent ways to solve equations of motion for 2D 
induced gravity. First, we find this solutions from the corresponding 
ones of the WZNW system, using the connection established in \cite{6}. 
We also solved 2D induced gravity equations of motion directly. In both 
cases we obtain the same result, given in (\ref{5.7}). It depends on two 
arbitrary fields $h^+$ and $h^-$, 
two arbitrary functions  $X_+(\eta_+)$ and $X_-(\eta_-)$, where 
$\eta_{\pm}$ are the solutions of the conditions $\pd_{\pm} \eta_{\mp}=0$, 
and three arbitrary constants $a,b$ i $c$. The metric tensor 
${\bar g}_{\m\n}$ has two independent components $h^+$ and $h^-$, because 
the corresponding scalar curvature is constant ${\bar R}= -\m$. 
Note that this is valid for every function $Y_+$ and $Y_-$.

As a consequence of the non-global nature of the Gauss decomposition \eq{4.1}, 
regular WZNW solutions can generate singular 2D gravity solutions. 
It was shown in ref. \cite{4} that the singularities of the Liouville 
variables are only coordinate singularities associated with the patching 
of $SL(2,R)$. 

The equations of motion for matter fields in only one sector have been 
considered before in the flat space-time. In this paper we generalized 
it to curved space-time. The solution for WZNW system was done in 
\eq{3.3} and \eq{3.4}, or in the absence of external fields in \eq{3.11}, 
while for 2D induced gravity in  \eq{5.6}. For flat space-time,  
$h^{\pm} \to \mp 1$, these equations go to the well known solutions 
of the WZNW model \eq{3.12} and of the Liouville theory \eq{5.6}, but in 
this case with $X_{\pm} = X_{\pm}(\xi^0 \pm \xi^1)$.     

Let us emphasis some advantage of the WZNW variables. 
The invariance of the solutions \eq{5.6} and (\ref{5.7}) under $SL(2,R)$ 
transformations is easy to find using the WZNW variables. The 
solution \eq{3.8} is invariant under the following transformations 
\be
G'_-=G_- k^{-1} \, , \qquad  G'_+= k G_+ \, , \qquad  e' =kek^{-1} \, , \leq{7.1}
\ee
where $k=const$ and  $k \in SL(2,R)$. 
If we take  
\be
k= \pmatrix { u  &  v   \cr 
              s  &  t   \cr } \,  ,                            \leq{7.2}
\ee
then the solutions \eq{5.6} as well as (\ref{5.7}) are invariant under 
\be
X'_+={ u X_+ + v \over s X_+ +t} \, , \qquad 
X'_-={ t X_- + s \over v X_- + u} \, ,                        \leq{7.3}
\ee
\be
Y'_+={ (au+cv ) Y_+ + v \over s Y_+ +a t-b s} \, , \qquad 
Y'_-={ (at-b s) Y_- + s \over v Y_- +au +cv} \, ,             \leq{7.4}
\ee
and $a'=aut -bus +cvt -dvs$ .

At the end we want to stress the advantage of using the $h^{\pm}$ components 
of the metric tensor, instead to fix them and go to the conformal gauge. 
These components gives us the opportunity to include topologically 
nontrivial solutions. It is known that in the three dimensional space of 
the metric tensor components (for two dimensional space-time) the 
region of the Minkowski signature have the group of integers as the 
first homotopy group \cite{10}. For each homotopy class $m$ we can choose 
metric configuration 
\be
{\bar g}_{\m\n}^m =e^{\rho} 
\pmatrix {\cos m\th   &  \sin m\th   \cr 
          \sin m\th   &  -\cos m\th   \cr }\, ,  \qquad   
         (\th=2 \arctan \s )                                 \leq{7.5}
\ee
representing noncontractible loops wound $m$ times around the forbidden 
region $det {\bar g}_{\m\n}=0$, when $\s$ goes from $-\infty$ to 
$\infty$. In our notation, for a kink number $m$, we have 
\be
e^{\tf_2} =e^{\rho} \cos m\th \,  , \qquad 
h^{\pm}= \mp {1 \mp \tan \fr{m\th}{2} \over 1 \pm \tan \fr{m\th}{2}}\, ,
                                                         \leq{7.6}
\ee
or 
\be
{\bar g}_{\m\n}^m ={\a^2 \over M}{\pd_+ Y_+ \pd_- Y_- \over (1- Y_+ Y_-)^2 } 
 \pmatrix {   1         &   \tan m\th   \cr 
             \tan m\th    &  -1          \cr }\, .     \label{7.7}
\ee
The usual gauge choice $h^{\pm}=\mp 1$ is too restrictive and in this case 
we lose all kink metrics.

\end{document}